\documentclass[12pt,a4paper]{article}
\usepackage[latin1]{inputenc}
\usepackage{amsmath}
\usepackage{amsmath} 
\usepackage{amsfonts}
\usepackage{amssymb}
\usepackage{float}
\usepackage{cite}
\usepackage{breqn}
\usepackage{graphics}
\usepackage{mathtools}
\usepackage{xcolor}

\usepackage{mathrsfs}

\newcommand{\plainfootnote}[1]{%
	\begingroup
	\renewcommand{\thefootnote}{}% Remove the footnote number
	\footnotetext{#1}%
	\addtocounter{footnote}{-1}% Prevent footnote counter from advancing
	\endgroup
}

\usepackage{fontawesome}
\usepackage{booktabs}

\definecolor{lightred}{rgb}{1, 0.5, 0.5}
\usepackage{tikz}
\usepackage{adjustbox}

\DeclareGraphicsExtensions{.pdf,.png,.jpg}
\usepackage[colorlinks=true,linkcolor=blue,citecolor=red]{hyperref}

\makeatletter
\newcommand{\mathleft}{\@fleqntrue\@mathmargin0pt}
\newcommand{\mathcenter}{\@fleqnfalse}
\makeatother
\newcommand*{\SavedEqref}{}
\let\SavedEqref\eqref
\renewcommand*{\eqref}[1]{%
	\begingroup
	\hypersetup{
		linkcolor=linkequation,
		linkbordercolor=linkequation,
	}%
	\SavedEqref{#1}%
	\endgroup
}

\usepackage[a4paper, width = 160mm, top = 30mm, bottom = 25mm, 
bindingoffset = 0mm]{geometry}

\def\beq{\begin{equation}}
	\def\eeq{\end{equation}}
\def\bea{\begin{eqnarray}}
	\def\eea{\end{eqnarray}}

\definecolor{lightorange}{RGB}{255, 178, 102} % Light Orange
\definecolor{mediumlightblue}{RGB}{100, 170, 255}   % Medium Light Blue
\definecolor{mediumlightred}{RGB}{230, 80, 80}      % Medium Light Red

\begin{document}
	
\begin{center}
	{\bf Tunneling time in non-Hermitian space fractional quantum mechanics
	}
	\vspace{0.7cm}

{\sf \small Mohammad Umar\textsuperscript{\textcolor{red}{\textnormal{1}}}, Vibhav Narayan Singh\textsuperscript{\textcolor{blue}{\textnormal{2}}}, Bhabani Prasad Mandal\textsuperscript{\textcolor{orange}{\textnormal{3}}}}

\bigskip
\plainfootnote{\textcolor{mediumlightred}{\faEnvelope}\textsuperscript{\textcolor{red}{\textnormal{}}} aliphysics110@gmail.com, opz238433@opc.iitd.ac.in}

\plainfootnote{\textcolor{mediumlightblue}{\faEnvelope}\textsuperscript{\textcolor{blue}{\textnormal{}}} vibhav.ecc123@gmail.com, vibhav.narayan@iilm.edu}

\plainfootnote{\textcolor{lightorange}{\faEnvelope}\textsuperscript{\textcolor{orange}{\textnormal{}}} bhabani.mandal@gmail.com, bhabani@bhu.ac.in}

{\em
	\textsuperscript{\textcolor{red}{\textnormal{1}}}Optics and Photonics Centre\\  
	Indian Institute of Technology Delhi\\ 
	New Delhi 110016, INDIA\\}
\vspace{0.5em}
{\em
	\textsuperscript{\textcolor{blue}{\textnormal{2}}}School of Basic \& Applied Sciences\\  
	IILM University\\ 
	Greater Noida 201308, INDIA\\
}
\vspace{0.5em}
{\em
	\textsuperscript{\textcolor{orange}{\textnormal{3}}}Department of Physics\\
	Banaras Hindu University\\  
	Varanasi 221005, INDIA\\}
\vspace{0.5em}

%%%%%%%%%%%%%%%%%%%%%%%%%%%%%%%%%%%%%%% Abstract %%%%%%%%%%%%%%%%%%%%%%%%%%%%%%%%%%%%

\vspace{0.8cm}	
\noindent {\bf Abstract}		
\end{center}

\noindent
We investigate the tunneling time of a wave packet propagating through a non-Hermitian potential $V_{r} - iV_{i}$ in space-fractional quantum mechanics. By applying the stationary phase method, we derive a closed-form expression for the tunneling time for this system. This study presents the first investigation of tunneling time at the interplay of non-Hermitian quantum mechanics and space-fractional quantum mechanics. The variation in tunneling time as the system transitions from a real to a complex potential is analyzed. We demonstrate that the tunneling time exhibits a dependence on the barrier width $d$ in the limit $d\rightarrow \infty$, showing the absence of the Hartman effect. A particularly striking feature of our findings is the potential manifestation of the Hartman effect for a specific combination of the absorption component $V_{i}$ and the Levy index $\alpha$. This behavior arises from the fact that the presence of the absorption component $V_{i}$ leads to a monotonic increase in tunneling time with barrier thickness, whereas the Levy index $\alpha$ reduces the tunneling time. The interplay of these contrasting influences facilitates the emergence of the Hartman effect under a specific combination of $V_{i}$ and the fractional parameter $\alpha$.
\medskip
\vspace{1in}
\newpage
%
%
%%%%%%%%%%%%%%%%%%%%%%%%%%%%%%%%% Introduction %%%%%%%%%%%%%%%%%%%%%%%%%%%%%%%%%%%%
%
%
\section{Introduction}
Hermiticity of a Hamiltonian is a fundamental postulate in quantum mechanics (QM), ensuring real eigenvalues and the conservation of probability through unitary time evolution in the standard Hilbert space. However, over the last two decades, a certain class of non-Hermitian systems with real energy eigenvalues has emerged as a frontier area of research, demonstrating that fully consistent quantum theories can be formulated by restoring an equivalent Hermiticity and preserving unitary dynamics in a modified Hilbert space \cite{bender1998real, bender2007making, mostafazadeh2006metric, mostafazadeh2010pseudo}. 
The theoretical predictions of non-Hermitian quantum mechanics (NHQM) have been rigorously validated through experimental implementations, particularly in the domain of optics \cite{musslimani2008optical, ruter2010observation, el2007theory, guo2009observation}. The experimental realization of non-Hermitian systems in laboratory settings has stimulated significant interest in both theoretical and experimental investigations \cite{berry2004physics, mostafazadeh2011spectral, mostafazadeh2012optical, mostafazadeh2009spectral, hajong2024hellmann, hasan2020new, ghatak2013various, kumari2016scattering, modak2021eigenstate, shukla2023uncertainty, mandal2015pt, mandal2005pseudo, mandal2012spectral, ghatak2019new, ghatak2012spectral, longhi2011invisibility, mostafazadeh2013invisibility,wu2022unidirectional, deak2012reciprocity, samsonov2011scattering, mostafazadeh2009resonance, longhi2010spectral, mostafazadeh2015physics, gmachl2010suckers, longhi2010backward, wan2011time, liu2010infrared, noh2012perfect, longhi2011coherent, hasan2014critical, heiss2004exceptional, heiss2012physics, ji2022tracking, cai2000observation, tischler2006critically, balci2011critical, hasan2020role}. Non-Hermitian Hamiltonians exhibit a plethora of unique scattering properties that fundamentally differ from those observed in Hermitian systems, introducing novel and previously unexplored phenomena. Notable features such as, invisibility \cite{longhi2011invisibility,mostafazadeh2013invisibility, wu2022unidirectional}, reciprocity \cite{deak2012reciprocity}, spectral singularities (SSs) \cite{mostafazadeh2011spectral, mostafazadeh2012optical, mostafazadeh2015physics, mostafazadeh2009spectral, samsonov2011scattering, mostafazadeh2009resonance, longhi2010spectral}, coherent perfect absorption (CPA) \cite{gmachl2010suckers, longhi2010backward, wan2011time, liu2010infrared, noh2012perfect, mostafazadeh2012optical, longhi2011coherent, hasan2014critical},  exceptional points (EPs) \cite{heiss2004exceptional, heiss2012physics, ji2022tracking} and critical coupling (CC) \cite{hasan2014critical, cai2000observation, tischler2006critically, balci2011critical} have garnered substantial attention due to their practical implications and pivotal role in advancing the understanding of diverse optical systems.\\
\indent
Soon after the extension of Hermitian QM to NHQM, a new generalization of QM based on its path integral (PI) formalism was also introduced. In the PI formulation of QM \cite{hibbs1965quantum}, PIs are evaluated over Brownian paths, which correspond to random processes governed by a Gaussian probability distribution and yield the Schrodinger equation. The Brownian process, however, is a specific case within a broader class of random processes known as Levy $\alpha$-stable random processes. Levy $\alpha$-stable random processes are non-Gaussian in nature and are characterized by the Levy index $\alpha$, where $0 < \alpha \leq 2$. Notably, when $\alpha = 2$, the Levy process reduces to the Brownian process, meaning that Levy paths are the Brownian paths for this specific case. Nick Laskin generalized the PI approach in QM by considering PIs over Levy paths \cite{laskin2000fractional,laskin20001fractional,laskin2002fractional}. The PI formulation over Levy paths resulted in the fractional Schrodinger equation, forming the foundation of a distinct branch of QM referred to as SFQM. After this, Naber \cite{naber2004time} formulated the time-fractional Schrodinger equation, while Wang and Xu \cite{wang2007generalized} extended it to a space-time fractional Schrodinger equation. \\
\indent
SFQM has garnered significant interest among researchers and has become a focus of extensive exploration by numerous researchers \cite{dong2007some, el2019some, kirichenko2018confinement, medina2019nonadiabatic, zhang2015propagation, ghalandari2019wave, yao2018solitons, xiao2018surface, zhang2017resonant, chen2018optical, wang2019elliptic, zhang2017unveiling, hasan2018new, singh2023quantum, khalili2021fractal, baleanu2009solving, golmankhaneh2024fractal} and various method have been employed in these studies such as adomain decomposition method (ADM) \cite{rida2008solution}, energy-conservative difference schemes \cite{wang2015energy}, conservative
finite element methods \cite{li2018fast}, fractional Fan sub-equation methods \cite{younis2017dark}, time-splitting Fourier pseudo-spectral method \cite{kirkpatrick2016fractional}, and transfer-matrix methods \cite{tare2014transmission}. SFQM is inspired by fractional calculus also called $F^{\alpha}$-calculus \cite{golmankhaneh2023fractal}, in which the derivative is of fractional order. This fractional approach is not only limited to quantum mechanics but also can be seen in plasma physics \cite{faridi2021fractional, abdelwahed2022physical}, optics \cite{ zhang2016diffraction}, astrophysics and cosmology \cite{ rasouli2024fractional, rasouli2022inflation, jalalzadeh2022sitter, el2013fractional, rami2015fractional, el2017fractional, el2017wormholes, el2013non, el2012gravitons, micolta2023revisiting, landim2021fractional, jalalzadeh2021prospecting, moniz2020fractional,  junior2025fractional, costa2023estimated, trivedi2024fractional, bidlan2025reconstructing}, high energy physics \cite{herrmann2008gauge} and quantum field theory \cite{tarasov2014fractional, calcagni2021quantum}. The fractional Schrodinger equation (FSE) has also been explored in optics \cite{longhi2015fractional} and the experimental realization of the FSE in the temporal domain has recently been demonstrated in the optical realm \cite{liu2023experimental}. The interplay between SFQM and NHQM has been explored for investigating SS behavior with Levy index $\alpha$ \cite{hasan2018new}. 
The study of tunneling time in QM has also been investigated in SFQM \cite{hasan2018tunneling, hasan2020tunneling}.\\
\indent
The advancements in NHQM and SFQM are among the latest developments in QM. However, one of the earliest and most fundamental problems in QM, quantum tunneling \cite{gurney1928wave, condon1931quantum, merzbacher2002early, razavy2013quantum}, still suffers from a paradox. The question of how much time a quantum particle takes to tunnel through a classically forbidden potential barrier has been a longstanding and contentious topic in quantum physics, sparking debate and research for decades \cite{bohm1989quantum, wigner1955lower, hartman1962tunneling, buttiker1982traversal, fletcher1985time, landauer1989barrier, steinberg1995much, olkhovsky1995more, chiao1997vi, winful2006tunneling, ghatak2015hartman, hasan2020role, hasan2020hartman,longhi2022non, guo2023tunneling, hasan2021general,  wu2024superluminality} and still this is an open problem both theoretically and experimentally. In 1962, Hartman used the SPM to investigate the tunneling dynamics of a wave packet propagating through the classically forbidden region in a metal-insulator-metal (MIM) configuration \cite{hartman1962tunneling}. His results revealed that the tunneling time remains invariant with respect to changes in the barrier thickness, a counterintuitive phenomenon now referred to as the Hartman Effect. This independence of the time delay from the thickness of the opaque barrier remains one of the most interesting and puzzling aspect of quantum tunneling. 
Subsequent independent research by Fletcher further validated this phenomenon, demonstrating that the tunneling time for evanescent waves saturates with increasing thickness of the opaque barrier \cite{fletcher1985time}. 
This paradox has motivated numerous researchers to propose alternative definitions of tunneling time to address the apparent inconsistency. For further details, interested readers are referred to \cite{winful2006tunneling}. The tunneling time has also been studied in double-barrier systems, where it was observed that the traversal time depends neither on the barrier widths nor on the separation between the barriers \cite{olkhovsky2002superluminal}. This result aligns with the \textit{general} Hartman effect, demonstrating that tunneling time in multi-barrier systems depends on neither the barrier thickness nor the inter-barrier separation \cite{olkhovsky2002superluminal}. For complex potentials involving elastic and inelastic channels, it was observed that tunneling time saturates with barrier thickness under conditions of weak absorption \cite{paul2007tunneling, ghatak2015hartman}. Additionally, it has been shown that the \textit {general(ized)} Hartman effect \cite{hasan2021general} exists for periodic and super-periodic potentials (SPP) \cite{hasan2018super} as well as for Cantor fractal potentials \cite{singh2025quantum, umar2023quantum, narayan2023tunneling, singh2023quant, mohammad2025polyadic}, which are the special cases of SPP. \\
\indent
The growing interest in NHQM and SFQM has ignited a deeper exploration of tunneling time, both in NHQM \cite{hasan2020role, hasan2020hartman, longhi2022non, guo2023tunneling} and in SFQM \cite{hasan2018tunneling, hasan2020tunneling}. The studies \cite{hasan2020role, hasan2020hartman} demonstrate the existence of the Hartman effect in $\mathcal{PT}$-symmetric systems, while the work \cite{longhi2022non} shows that $\mathcal{PT}$-symmetry is not a necessary requirement for the Hartman effect to occur. Furthermore, the works \cite{hasan2018tunneling, hasan2020tunneling} reveal that the Hartman effect is absent in SFQM. Motivated by these advancements, we investigate tunneling time by using the SPM in the unique blend of NHQM and SFQM. This blend is referred to as non-Hermitian space fractional quantum mechanics (NHSFQM). This blend is reported in \cite{hasan2018new}, where the red and blue shift of SS are shown to depend on the Levy index $\alpha$. To the best of our knowledge, this study presents the first investigation of tunneling time within the framework of NHSFQM. We have calculated the tunneling time by employing the SPM method for a non-Hermitian potential $V_{r}-iV_{i}$ in SFQM. Our analysis reveals the potential manifestation of the Hartmann effect for specific combinations of $V_{i}$ and $\alpha$, which we have substantiated through graphical representations. The observed phenomenon can be attributed to the competing effects of the imaginary component of the potential and the Levy index $\alpha$, on tunneling time. Specifically, $V_i$ induces a monotonic increase in tunneling time with increasing thickness of the potential barrier, whereas $\alpha$ contributes to its reduction. In the context of a complex barrier system within SFQM, these opposing influences can counteract each other for specific parameter values, resulting in a cancellation effect that may give rise to the Hartman effect.\\
\newline
\indent
This paper is structured as follows: In Section \ref{section2}, we present the SPM for evaluating tunneling time and discuss the Hartman effect in standard QM. Section \ref{section3} introduces the space fractional Schrodinger equation. In Section \ref{section4}, we compute the tunneling time for transmission through a complex potential in SFQM using the SPM and demonstrate the nonexistence of the Hartman effect in this regime. Additionally, we discuss the potential manifestation of the Hartman effect under specific conditions. Finally, the study reaches its conclusion in Section \ref{section5}.
\section{Tunneling time and the Hartman effect}
\label{section2}
In this section, we employ the stationary phase approximation to determine the tunneling time of a free particle across the classically forbidden region of space \cite{roy2009elements}. Within this framework, the tunneling time is defined as the temporal delay between the peak positions of the incident and transmitted localized wave packets as they traverse the potential barrier. To find the tunneling time $\Gamma$, consider the time evolution of a localized wave packet $\mathbb{G}_{k_0}(k)$, given by a normalized Gaussian function having a peak at the mean momentum $\hbar k_0$:
\begin{equation}
	\int \mathbb{G}_{k_0}(k) e^{i(kx - \frac{Et}{\hbar})} \, dk, \tag{1}
\end{equation}
where the wave number $k = \sqrt{2mE}$. The wave packet propagates in the positive $x$-direction. As a result of its interaction with the potential barrier, the transmitted wave packet is modified as follows:
\begin{equation}
	\int \mathbb{G}_{k_0}(k) |\mathscr{T}(k)| e^{i(kx - \frac{Et}{\hbar} + \Phi(k))} \, dk, \tag{2}
\end{equation}
where $\mathscr{T}(k) = |\mathscr{T}(k)| e^{i\Phi(k)}$ represents the transmission coefficient associated with the rectangular potential barrier $V(x)$, which is defined as $V(x) = V$ for $0 \leq x \leq d$, and zero elsewhere. According to the SPM, we have
\begin{equation}
	\frac{d}{dk} \left(kd - \frac{E \Gamma}{\hbar} + \Phi(k)\right) = 0. \tag{3}
\end{equation}
This gives the tunneling time expression as
\begin{equation}
	\Gamma = \hbar \frac{d\Phi(E)}{dE} + \frac{d}{\left(\frac{\hbar k}{m}\right)}. \tag{4}
\end{equation}
For a square barrier potential $V(x) = V$ confined to the region $0 \leq x \leq d$ and zero elsewhere, the tunneling time is given by:
\begin{equation}
	\Gamma = \hbar \frac{d}{dE}\left\{ \tan^{-1} \left(\frac{k^2 - \kappa^2}{2k\kappa} \tanh\kappa d\right) \right\}, \tag{5}
\end{equation}
where $\kappa = \sqrt{2m(V - E)}/\hbar$. In the limit $d \to 0$, we have $\Gamma \to 0$ as expected. However, for $d \to \infty$, the tunneling time becomes $\Gamma =2m/\hbar k \kappa$, demonstrating that it remains invariant with respect to the barrier width $d$ for sufficiently thick barriers. This phenomenon, known as the Hartman effect, states that the tunneling time remains constant irrespective of the barrier thickness when the barrier is sufficiently wide. In a unit system where $2m = 1$, $\hbar = 1$, and $c = 1$, the tunneling time simplifies to  
\begin{equation}
	\lim_{d \to \infty} \Gamma \sim \frac{1}{k\kappa}. \tag{7}
\end{equation}
\section{The fractional Schrodinger equation}
\label{section3}
The one-dimensional space fractional Schrodinger is expressed as \cite{laskin2000fractional,laskin20001fractional, laskin2002fractional} 
\begin{equation}
	i\hbar \frac{\partial \psi(x, t)}{\partial t} = \mathcal{H}_\alpha(x, t) \psi(x, t), \quad 1 < \alpha \leq 2,
	\label{frac_01}
\end{equation}
where $\mathcal{H}_\alpha(x, t)$ is the fractional Hamiltonian operator, expressed through the Riesz fractional derivative $(-\hbar^2 \Delta)^{\alpha/2}$ as
\begin{equation}
	\mathcal{H}_\alpha(x, t) = D_\alpha (-\hbar^2 \Delta)^{\alpha/2} + V(x, t).
\end{equation}
Here, $\Delta = \frac{\partial^2}{\partial x^2}$ and $D_{\alpha}$ is the generalized fractional quantum diffusion coefficient. The physical dimension of $D_{\alpha}$ is given by $[D_{\alpha}]=\texttt{erg}^{1-\alpha}\times \texttt{cm}^{\alpha}\times \texttt{s}^{-\alpha}$, which depends on the Levy index $\alpha$. For $\alpha=2$, $D_{\alpha}$ simplifies to $1/2m$, where $m$ is the mass of the particle. The Riesz fractional derivative of the wave function $\psi(x, t)$ is defined as
\begin{equation}
	(-\hbar^2 \Delta)^{\alpha/2} \psi(x, t) = \frac{1}{2\pi\hbar} \int_{-\infty}^\infty \widetilde{\psi}(p, t) |p|^\alpha e^{ipx/\hbar} \, dp,
\end{equation}
where $\widetilde{\psi}(p, t)$ is the Fourier transform of $\psi(x, t)$:
\begin{equation}
	\widetilde{\psi}(p, t) = \mathcal{F}[\psi(x, t)] = \int_{-\infty}^\infty \psi(x, t) e^{-ipx/\hbar} \, dx.
\end{equation}
This study focuses on the scenario where $V(x,t)=V(x)$, ensuring that the fractional Hamiltonian operator $\mathcal{H}_{\alpha}$ remains time-independent. Consequently the Eq. (\ref{frac_01}) takes the following form 
\begin{equation}
	D_\alpha (-\hbar^2 \Delta)^{\alpha/2} \psi(x) + V(x) \psi(x) = E \psi(x)
\end{equation}
after the separation of variables. Now, the time-dependent wave function $\psi(x,t)$ is given by $\psi(x,t) = \psi(x)e^{-iEt/\hbar}$, where $\psi(x)$ represents the time-independent wave function and $E$ corresponds to the energy of the quantum system. In this study, the generalized diffusion coefficient is taken as  
\begin{equation}
	D_{\alpha}=\frac{u^{2-\alpha}}{\alpha m^{\alpha-1}},
\end{equation}
where $u$ represents the characteristic velocity of the non-relativistic quantum system, taken as $1.0 \times 10^{-5}c$, with $c$ denoting the speed of light in vacuum.
\section{Tunneling time from a non-Hermitian potential in space fractional quantum mechanics}
\label{section4}
\begin{figure}[H]
	\begin{center}
		\includegraphics[scale=0.43]{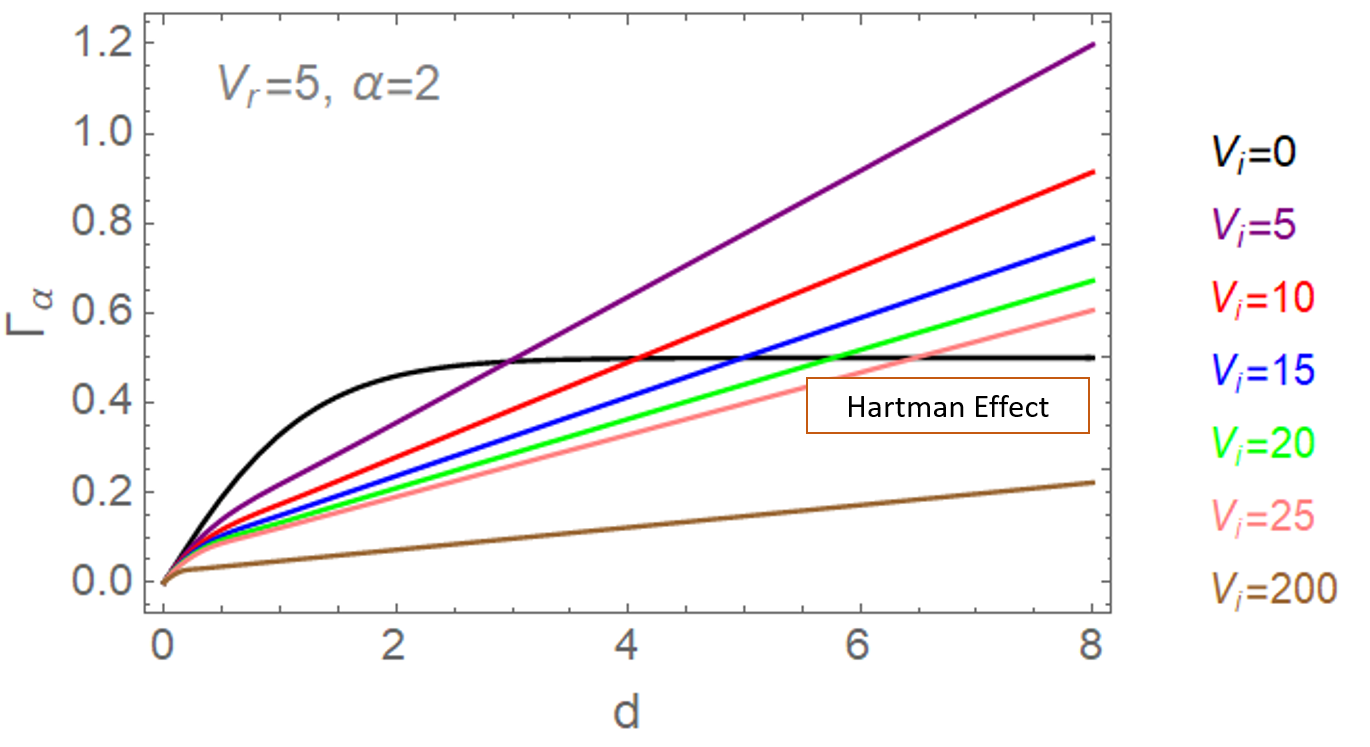}
		\caption{(Color online).  \it Figure presents the variation of tunneling time $\Gamma_{\alpha}$ as a function of barrier width $d$ for a complex potential in standard QM ($\alpha = 2$), with $V_r = 5$ and different values of $V_i$, for an incident energy of $E = 4$ ($< V_r$). The system exhibits the well-known Hartman effect for the real potential and this effect disappears when $V_{i}> 0$.
		}
		\label{tt_01}
	\end{center}
\end{figure}

\begin{figure}[H]
	\begin{center}
		\includegraphics[scale=0.382]{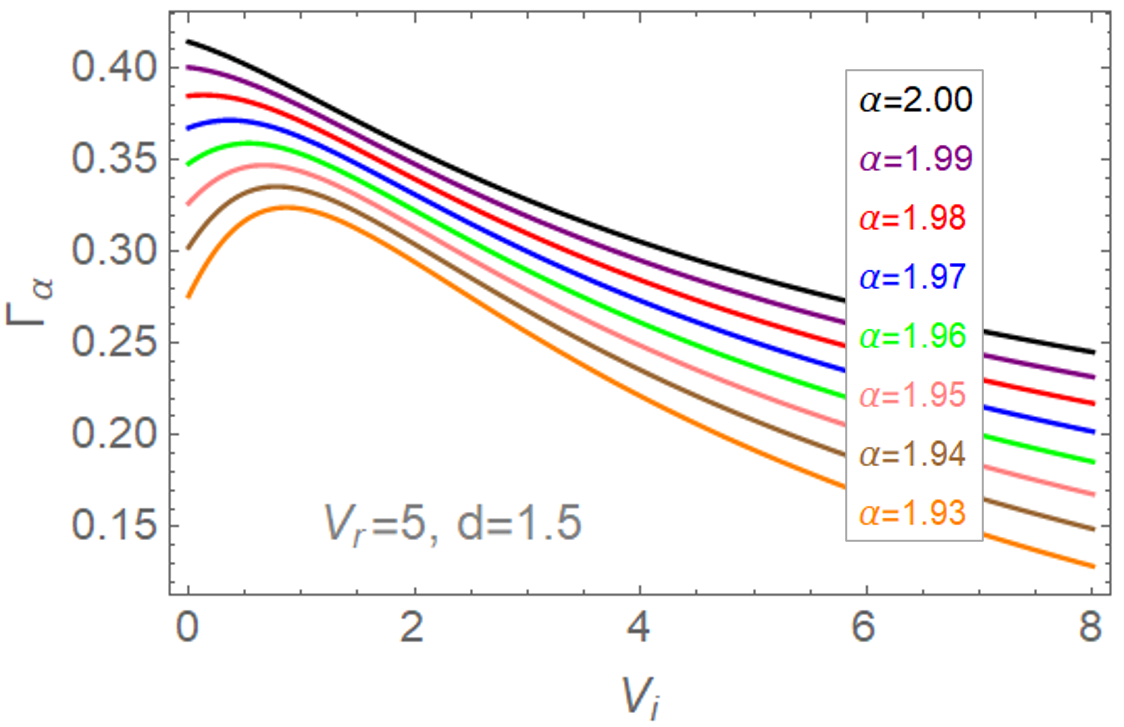} (a)
		\includegraphics[scale=0.382]{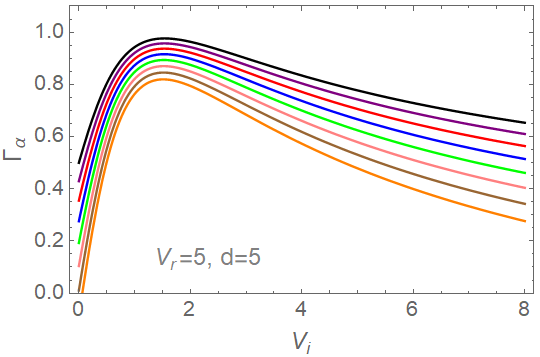} (b)
		\caption{(Color online). \it Figure presents the variation of tunneling time $\Gamma_{\alpha}$ with the absorption component $V_i$ for different values of Levy index $\alpha$ and for fixed barrier thicknesses (a) $d = 1.5$ and (b) $d = 5$. The incident wave energy is fixed at $E = 4$. For high value of $V_{i}$, the tunneling time decreases, and as $\alpha$ decreases, the tunneling time also decreases.
		}
		\label{tt_003}
	\end{center}
\end{figure}

\begin{figure}[H]
	\begin{center}
		\includegraphics[scale=0.31]{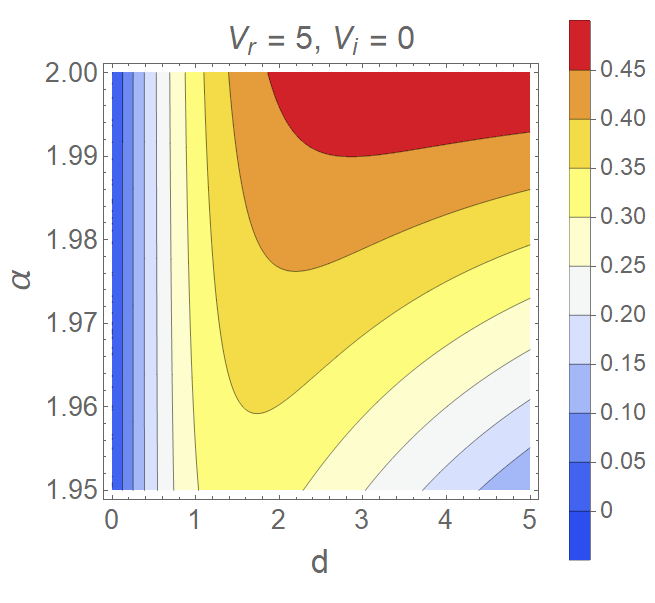} (a)
		\includegraphics[scale=0.31]{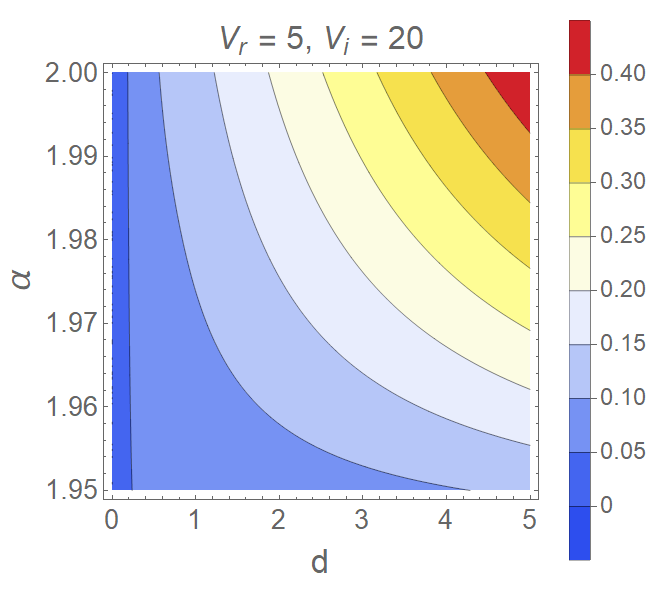} (b)
		\caption{(Color online). \it Contour plots illustrating the variation of tunneling time in the $\alpha-d$-plane for (a) a real potential ($V_{i}=0$) and for (b) a complex potential ($V_{i}=20$). The incident wave energy is fixed at $E=4$. The color scale represents the tunneling time.}
		\label{add_01}
	\end{center}
\end{figure}

Consider a complex potential $V=V_{r}-iV_{i}$ confined over the region $0\le x \le d$ and zero elsewhere, then the corresponding space-fractional Schrodinger equation is
\begin{equation}
	D_{\alpha}(-\hbar^{2}\Delta)^{\alpha/2}\psi(x)+V\psi(x)=E\psi(x).
	\label{s_01}
\end{equation}
The general solution of Eq. (\ref{s_01}) has the following form:
\begin{align}
	\psi(x) =
	\begin{cases}
		Ae^{i k_\alpha x} + Be^{-i k_\alpha x}, & x < 0, \\
		C \cos \kappa_\alpha x + D \sin \kappa_\alpha x, & 0 < x < d, \\
		Fe^{i k_\alpha x} + Ge^{-i k_\alpha x}, & x > d,
	\end{cases} \tag{14}
\end{align}
where
\begin{equation}
	k_{\alpha} = \left( \frac{E}{D_\alpha \hbar^\alpha} \right)^{\frac{1}{\alpha}},
\end{equation}
\begin{equation}
	\kappa_{\alpha} = \left( \frac{E - V_{r}+iV_{i}}{D_\alpha \hbar^\alpha} \right)^{\frac{1}{\alpha}}.
\end{equation}
The transmission coefficient for a particle traversing a rectangular barrier in SFQM has been determined in \cite{guo2006some}, with the corresponding transfer matrix formulation provided in \cite{tare2014transmission}. The transmission coefficient for a rectangular potential barrier in SFQM is given by \cite{hasan2018tunneling}
\begin{equation}
	T(k_{\alpha}, \kappa_{\alpha}) = \frac{e^{-i k_\alpha d}}{\cos \kappa_\alpha d - i \mu \sin \kappa_\alpha d}, 
\end{equation}
where
\begin{equation}
	\mu = \frac{\rho + \rho^{-1}}{2} \quad \text{and} \quad 
	\rho = \left( \frac{k_\alpha}{\kappa_\alpha} \right)^{\alpha - 1}.
\end{equation}
For the classically forbidden case ($E<V_{r}$), the expression for transmission coefficient can be expressed as 
\begin{equation}
	\mathcal{T}(k_{\alpha}, \widetilde{\kappa_\alpha}) = \frac{e^{-i k_\alpha d}}{\cos \widetilde{\kappa_\alpha} d - i \widetilde{\mu} \sin \widetilde{\kappa_\alpha} d},
	\label{transmission_01}
\end{equation}
where
\begin{align}
	\widetilde{\kappa_\alpha} &= (-1)^{\frac{1}{\alpha}} \left( \frac{V_{r} - E - iV_{i}}{D_\alpha \hbar^\alpha} \right)^{\frac{1}{\alpha}}= e^{i\frac{\pi}{\alpha}} (\chi + i\eta).
	\label{kappa_alpha}
\end{align}
In the above expression, $\chi$ and $\eta$ is expressed as 
\begin{equation}
	\chi = \left(\frac{\sqrt{U}}{D_{\alpha}}\right)^{\frac{1}{\alpha}} \cos\frac{\theta}{\alpha} \quad \text{and} \quad
	\eta = \left(\frac{\sqrt{U}}{D_{\alpha}}\right)^{\frac{1}{\alpha}} \sin\frac{\theta}{\alpha}.
\end{equation}
Here, $U = (V_{r} - E)^{2} + V_{i}^{2}$ and $\theta = -\tan^{-1}\left(\frac{V_{i}}{V_{r} - E}\right)$.
In Eq. (\ref{transmission_01}), $\widetilde{\mu}$ is expressed as \begin{equation}
	\widetilde{\mu} = \frac{1}{2}\left[\left( \frac{k_\alpha}{\widetilde{\kappa_\alpha}} \right)^{\alpha - 1}+\left( \frac{\widetilde{\kappa_\alpha}}{k_\alpha} \right)^{\alpha - 1}\right] = \mu_{1}+i\mu_{2},
\end{equation}
where, $\mu_{1}$ and $\mu_{2}$ is given by  
\begin{equation}
	\mu_{1} = \frac{1}{2}\left[\left( \frac{k_\alpha}{\sqrt{\gamma}} \right)^{\alpha - 1}+\left( \frac{\sqrt{\gamma}}{k_\alpha} \right)^{\alpha - 1}\right]\cos[\lambda(\alpha-1)],
	\label{15}
\end{equation}
\begin{equation}
	\mu_{2} = \frac{1}{2}\left[\left( \frac{k_\alpha}{\sqrt{\gamma}} \right)^{\alpha - 1}-\left( \frac{\sqrt{\gamma}}{k_\alpha} \right)^{\alpha - 1}\right]\sin[\lambda(\alpha-1)].
	\label{16}
\end{equation}
In the above expressions, $\gamma=\chi^{2}+\eta^{2}$ and $\lambda$ is given by 
\begin{equation}
	\lambda =-\frac{\pi}{\alpha}-\tan^{-1}\left(\frac{\eta}{\chi}\right).
\end{equation}
The expression for transmission coefficient, Eq. (\ref{transmission_01}), can be further expressed as with it denominator in complex form as
\begin{equation}
	\mathcal{T}(k_{\alpha}, \widetilde{\kappa_\alpha}) = \frac{e^{-i k_\alpha d}}{\xi-i\zeta},
	\label{tcomplex}
\end{equation}
where $\xi = \texttt{Re}[\cos \widetilde{\kappa_\alpha} d - i \widetilde{\mu} \sin \widetilde{\kappa_\alpha} d]$ and $\zeta = \texttt{Im}[\cos \widetilde{\kappa_\alpha} d - i \widetilde{\mu} \sin \widetilde{\kappa_\alpha} d]$ and expressed as 
\begin{equation}
	\xi = \cos\lambda_{1}d(\cosh\lambda_{2}d+\mu_{1}\sinh\lambda_{2}d)+\mu_{2}\sin\lambda_{1}d\cosh\lambda_{2}d,
	\label{xi}
\end{equation} 
\begin{equation}
	\zeta =\sin\lambda_{1}d(\sinh\lambda_{2}d+\mu_{1}\cosh\lambda_{2}d)-\mu_{2}\cos\lambda_{1}d\sinh\lambda_{2}d.
	\label{zeta}
\end{equation}
In the above expressions, $\lambda_{1}$ and $\lambda_{2}$ is expressed as 
\begin{align}
	\lambda_{1} &= \chi \cos{\frac{\pi}{\alpha}} - \eta \sin{\frac{\pi}{\alpha}}, \\
	\lambda_{2} &= \eta \cos{\frac{\pi}{\alpha}} + \chi \sin{\frac{\pi}{\alpha}}.
\end{align}
From Eq. (\ref{tcomplex}), the net tunneling phase can be written as
\begin{equation}
	\Phi_{\alpha}=\phi-k_{\alpha}d,
\end{equation}
where $\phi = \tan^{-1}(\zeta/\xi)$.
Therefore, the tunneling time through a complex potential in SFQM is expressed as
\begin{align}
	\Gamma_{\alpha} 
	&= \frac{J_{\alpha}}{H_{\alpha}}-\frac{dk_{\alpha}^{1-\alpha}}{\alpha D_{\alpha}}+\frac{d}{\frac{\hbar k}{m}},
	\label{gamma}
\end{align}
where $J_{\alpha}$ and $H_{\alpha}$ is given by
\begin{multline}
	J_{\alpha} = \frac{1}{2}\Big[(\mu_{1}\mu_{2}^{'}-\mu_{2}(\mu_{1}^{'}+2d\lambda_{2}^{'}))\cos2d\lambda_{1} +(\mu_{1}^{'}\mu_{2}-\mu_{1}\mu_{2}^{'}+2d\mu_{1}\lambda_{1}^{'})\cosh2d\lambda_{2}\\+(\mu_{1}^{'}-d\lambda_{2}^{'}(\mu_{1}^{2}+\mu_{2}^{2}-1)\lambda_{2}^{'})\sin2d\lambda_{1}+(-\mu_{2}^{'}+d\lambda_{1}^{'}(\mu_{1}^{2}+\mu_{2}^{2}+1))\sinh2d\lambda_{2}\Big],
	\label{j_alpha}
\end{multline}
\begin{multline}
	H_{\alpha}=\frac{1}{2}\Big[(\mu_{1}^{2}+\mu_{2}^{2}+1)\cosh2d\lambda_{2}-(\mu_{1}^{2}+\mu_{2}^{2}-1)\cos2d\lambda_{1}\Big]+\mu_{2}\sin2d\lambda_{1}+\mu_{1}\sinh2d\lambda_{2}.
\end{multline}
In Eq. (\ref{j_alpha}), the prime notation denotes differentiation with respect to $E$, i.e., $\mu_{1,2}^{'} = d\mu_{1,2}/dE$ and $\lambda_{1,2}^{'} = d\lambda_{1,2}/dE$. Next, we investigate the tunneling time in the limit  $d\rightarrow \infty$ to evaluate the possibility of the Hartman effect. It is well known that 
\begin{equation}
	\lim_{x \to \infty} \sinh x \sim \frac{1}{2} e^{x} \sim \lim_{x \to \infty} \cosh x.
\end{equation}
By using the above fact, we simplify
\begin{equation}
	\lim_{d \to \infty} J_{\alpha} \sim \frac{e^{2d\lambda_2}}{4} \Big[d\lambda_{1}^{'} (2\mu_1 + \mu_1^2 + \mu_2^2 + 1)+[\mu_{1}^{'}\mu_{2}-\mu_{2}^{'}(\mu_{1}+1)]\Big],
\end{equation}
\begin{equation}
	\lim_{d \to \infty} H_{\alpha}\sim \frac{e^{2d\lambda_2}}{4} (2\mu_1 + \mu_1^2 + \mu_2^2 + 1).
\end{equation}
In this limiting case, the tunneling time (Eq. \ref{gamma}) is expressed as
\begin{equation}
	\lim_{d \to \infty} \Gamma_{\alpha}\sim d\left(\lambda_{1}^{'}-\frac{k_{\alpha}^{1-\alpha}}{\alpha D_{\alpha}}+\frac{1}{2k}\right)+[\mu_{1}^{'}\mu_{2}-\mu_{2}^{'}(\mu_{1}+1)].
	\label{eq30}
\end{equation}
This shows that in the limit $d \rightarrow \infty$, the tunneling time exhibits a dependence on the width 
$d$, showing the absence of the Hartman effect in this scenario. Fig. \ref{tt_01} presents the variation of tunneling time as a function of barrier width for a complex potential in standard QM ($\alpha = 2$), considering $V_{r}=5$ and different values of $V_{i}$, with an incident energy of $E = 4$ ($< V_{r}$). The results indicate that, in the absence of an imaginary (lossy) component ($V_{i} = 0$), the system exhibits the well-known Hartman effect, as depicted by the black curve. Further, when the potential contains an imaginary component, the Hartman effect does not exist. In the presence of a complex potential, the tunneling time increases with the barrier width up to a certain threshold, beyond which its rate of increase becomes comparatively slower. Another observation is that, as the lossy component of the potential, $V_{i}$ ($> 0$), increases, the tunneling time exhibits a systematic reduction. This behavior arises due to enhanced absorption within the potential barrier, which suppresses the wavefunction amplitude and consequently diminishes the phase accumulation necessary for tunneling. As a result, the SPM yields a shorter tunneling time, indicating that the dominant tunneling pathways are those experiencing minimal attenuation. Overall, for a given $V_{i}$, the tunneling time increases monotonically with the barrier thickness, while it decreases as $V_{i}$ increases. Fig. \ref{tt_003}a and \ref{tt_003}b presents the variation of tunneling time as a function of the lossy component $V_{i}$ for barrier widths $d=1.5$ and $d=5$ respectively. Fig.~\ref{add_01} provides a detailed visualization of the tunneling time through contour plots, depicting its dependence on the parameters in the $\alpha-d$ plane. Fig. \ref{add_01}a corresponds to the case of a real potential ($V_i = 0$), while Fig. \ref{add_01}b illustrates the effect of a complex potential ($V_i = 20$). The color variation represents the tunneling time. Fig.~\ref{add_01} provides a detailed visualization of the tunneling time through contour plots, depicting its dependence on the parameters in the $\alpha-d$ plane. Fig. \ref{add_01}a corresponds to the case of a real potential ($V_i = 0$), while Fig. \ref{add_01}b illustrates the effect of a complex potential ($V_i = 20$). The color variation represents the tunneling time.
\noindent

\begin{figure}[H]
	\begin{center}
		\includegraphics[scale=0.45]{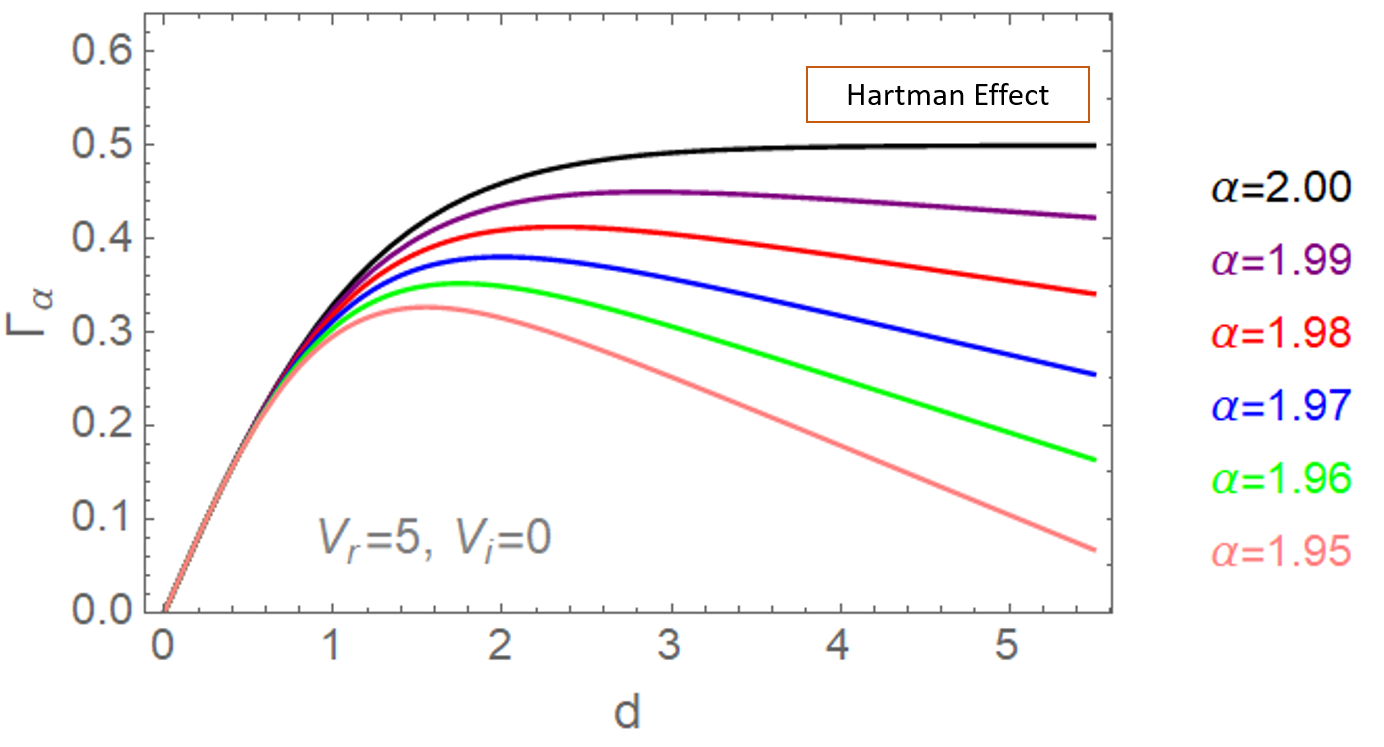}
		\caption{(Color online). \it Variation of tunneling time $\Gamma_{\alpha}$ with barrier thickness $d$ for different values of $\alpha$. For $\alpha<2$, tunneling time initially rises, reaching a peak at a specific threshold $d$ and then begins to decrease with increasing $d$. For $\alpha=2$, the well-known Hartman effect is recovered. The incident wave energy is fixed at $E = 4$.}
		\label{tt_02}
	\end{center}
\end{figure}

\begin{figure}[H]
	\begin{center}
		\includegraphics[scale=0.321]{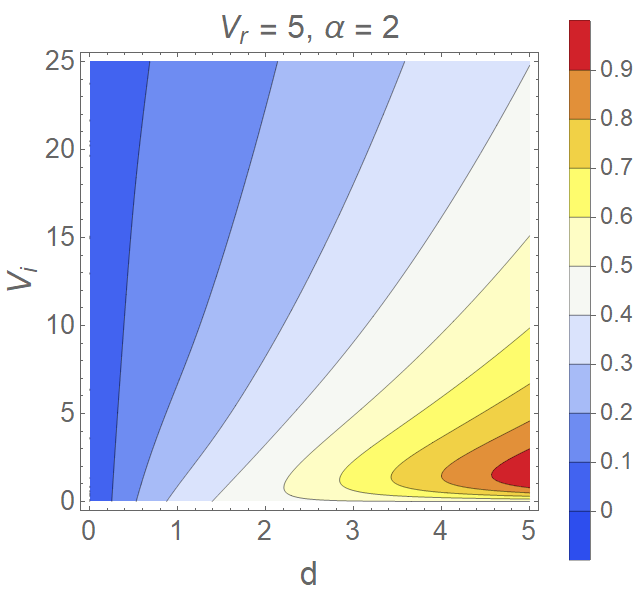} (a)
		\includegraphics[scale=0.321]{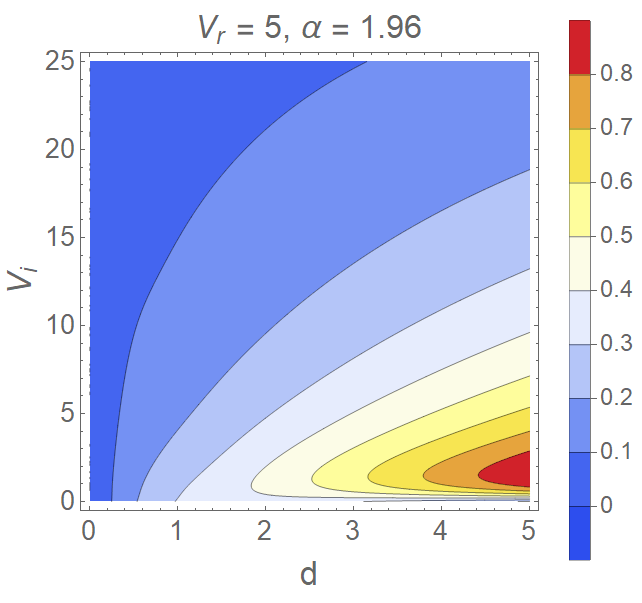} (b)
		\caption{(Color online). \it Contour plots illustrating the variation of tunneling time in the $V_{i}-d$ plane, in (a) standard QM ($\alpha = 2$) and in (b) SFQM ($\alpha = 1.96$). The incident wave energy is fixed at $E = 4$. The color variation represents the tunneling time.
		}
		\label{vidd_01}
	\end{center}
\end{figure}

\begin{figure}[H]
	\begin{center}
		\includegraphics[scale=0.382]{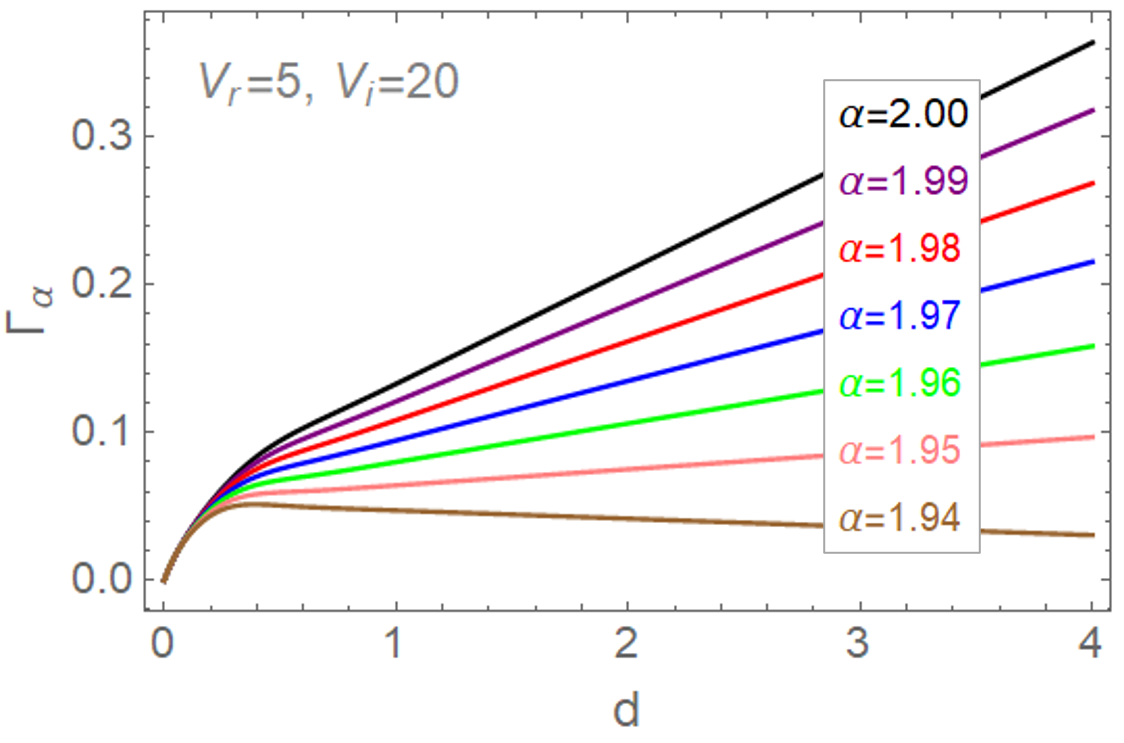} (a)
		\includegraphics[scale=0.382]{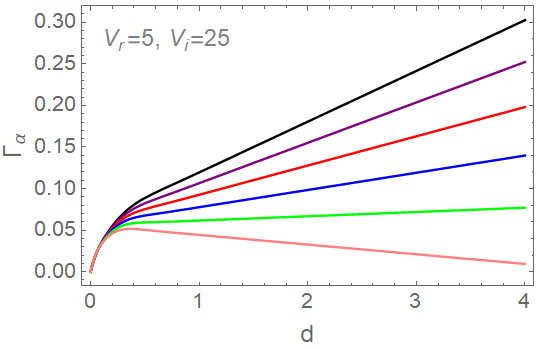} (b) \\
		\includegraphics[scale=0.382]{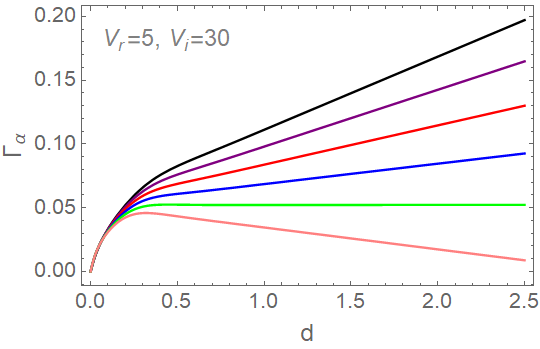} (c)
		\includegraphics[scale=0.382]{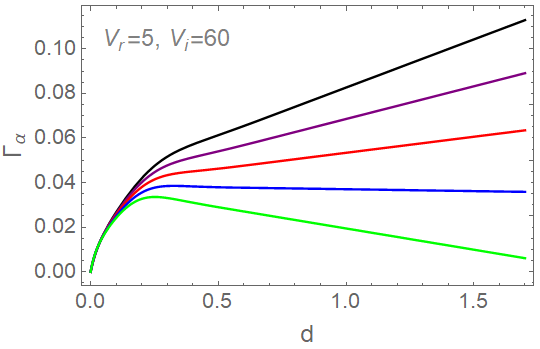} (d)
		\caption{(Color online). \it Variation of tunneling time as a function of barrier thickness $d$ for different values of the Levy index $\alpha$, considering various absorption component values as (a) $V_{i} = 20$, (b) $V_{i} = 25$, (c) $V_{i} = 30$ and (d) $V_{i} = 60$. The incident wave energy is fixed at $E = 4$. For a given $V_{i}$, the tunneling time decreases for $\alpha<2$. These plots also reveal a potential manifestation of the Hartman effect for a given combination of $V_{i}$ and fractional parameter $\alpha=\alpha_{H}$).}
		\label{tt}
	\end{center}
\end{figure}

Fig. \ref{tt_02} depict the variation of tunneling time with barrier width $d$ for the real potential ($V_{i}=0$) and for different values of Levy index $\alpha$. This result is in full concordance with the findings presented in \cite{hasan2018tunneling}. It is evident from these plots that in SFQM ($\alpha<2$), a particle requires less time to traverse the classically forbidden region compared to standard ($\alpha=2$). As the parameter $\alpha$ decreases, the tunneling time exhibits a corresponding decrease. Moreover, it is noteworthy that the Hartman effect is absent in SFQM. Beyond a certain threshold in the width of a classically opaque barrier, the tunneling time in SFQM exhibits a monotonically decreasing behavior with increasing barrier thickness; this behavior warrants further investigation.
\indent
Now, we turn to the study of tunneling time through a complex potential in SFQM, which forms the core objective of this paper. Fig. \ref{vidd_01} offers a detailed comparative visualization of the tunneling time from a complex potential in standard QM ($\alpha=2$) and in SFQM ($\alpha=1.96$) through the contour plots in the $V_{i}-d$ plane. Next, Fig. \ref{tt} presents the tunneling time behavior for different values of the absorption component $V_i$ of the complex potential, and the fractional parameter $\alpha$. Specifically, Fig. \ref{tt}(a) corresponds to the potential $5 - 20i$. For $\alpha = 2$, as previously discussed, the Hartman effect is absent. In this case, the tunneling time initially increases with the barrier width up to a certain threshold, beyond which it continues to increase linearly but at a comparatively slower rate. A similar trend is seen for $\alpha < 2$ up to a specific threshold, say as $\alpha = \alpha_H$. Beyond
\noindent
this threshold, the tunneling time first increases with the barrier width up to a certain value of $d$, after which it starts to decrease as the barrier thickness increases.\\
\indent
It is important to note that the tunneling time decreases as $\alpha$ decreases, similar to the behavior observed in the real potential case (Fig. \ref{tt_02}). Additionally, at $\alpha_{H}$ there is a potential manifestation of the Hartman effect. In Fig. \ref{tt}a, this $\alpha_{H}$ lies in the range of $1.95<\alpha <1.94$ as evident from the plot. As I discussed earlier, (Fig. \ref{tt_01}) as the absorption component $V_{i}$ increases, the tunneling time decreases. This trend is further corroborated in Fig. \ref{tt}. Now, consider Fig. \ref{tt}b which corresponds to the potential $5 - 25i$, where it is also observed that the Hartman effect can manifest at $\alpha_{H}$ which lies around $\alpha = 1.96$. A similar pattern is observed in Fig. \ref{tt}c and Fig. \ref{tt}d, which correspond to the potentials $5 - 30i$ and $5 - 60i$, respectively. In Fig. \ref{tt}d, the possibility of the Hartman effect is identified around $\alpha=1.97$. These observations show that the presence of $V_{i}$ suppresses the Hartman effect, and this suppression also occurs in SFQM due to the fractional effect. In the first case, the Hartman effect is restored when $V_{i} = 0$, while in the second case, it reappears when $\alpha = 2$. The tunneling time through the complex potential in SFQM does not exhibit the Hartman effect, as analytically demonstrated in Eq. (\ref{eq30}). But, graphical analysis (Fig. \ref{tt}) reveals a very striking feature, suggesting that the Hartman effect can emerge for specific combinations of $(V_{i}, \alpha_{H})$. Here, $\alpha_{H}$ denotes the particular value of $\alpha$ that, for a given $V_{i}$, leads to the emergence of the Hartman effect.

\indent
The observed phenomenon is attributed to the interplay between the absorption component $V_i$ of the potential, and the fractional parameter $\alpha$, in modulating the tunneling time as a function of barrier thickness. Specifically, $V_i$ induces a monotonic increase in tunneling time with increasing barrier thickness $d$, consistent with the expected attenuation effects associated with absorption. Conversely, the fractional parameter $\alpha$ dictates the nonlocal properties of the transport dynamics, leading to a reduction in tunneling time for both real and complex potentials as its value decreases. In SFQM, the path integral formulation extends beyond Brownian trajectories to include Levy flight paths, which exhibit a higher probability of long-range jumps compared to conventional Brownian motion. This fundamental distinction leads to significant modifications in quantum transport properties. Specifically, due to the presence of Levy flights, a quantum particle in SFQM is more likely to traverse distant points within a single step, thereby reducing the effective traversal time across a classically forbidden region compared to standard QM. Therefore, lower values of $\alpha$ enhance the probability of long-range jumps, thereby further reducing the tunneling time. These contrasting effects (effect of $V_{i}$ and $\alpha$), introduce a nontrivial interplay, wherein the absorption-induced delay and the fractional-order suppression of tunneling time can counterbalance each other under specific parametric conditions. This intricate cancellation mechanism can, at specific values of $V_{i}$ and $\alpha = \alpha_{H}$, give rise to an effective restoration of the Hartman effect, despite the presence of absorption, a phenomenon that is otherwise conventionally associated with its suppression. This result is particularly striking and warrants further investigation.
\section{Conclusion}
\label{section5}
This study presents the first investigation of tunneling time through a non-Hermitian potential in SQFM, thereby bridging two fundamental domains of QM: NHQM and SFQM. This synthesis leads to the development of a new paradigm, which we refer to as non-Hermitian space-fractional quantum mechanics (NHSFQM). The SPM is employed to derive the tunneling time expression for the complex potential in SFQM. We investigate the influence of the absorption component $V_{i}$ and the Levy index $\alpha$ on the tunneling time. The presence of the absorption component $V_{i}$ suppresses the Hartman effect, leading to a monotonic increase in tunneling time with barrier thickness $d$. Additionally, as $V_{i}$ increases, the tunneling time decreases. The Hartman effect is also diminished in the fractional regime, with a reduction observed for $\alpha < 2$ in both real and complex potentials. A particularly intriguing observation is that while both $V_{i}$ and $\alpha$ suppress the Hartman effect, a specific combination of $V_{i}$ and $\alpha = \alpha_{H}$ may lead to its manifestation. This phenomenon is of significant interest and necessitates further investigation.

\indent
It is important to highlight that in the context of the space fractional Schrodinger equation (SFSE), the fractional parameter $\alpha$ takes values within the range $0 < \alpha \leq 2$. Specifically, our study focuses on examining the tunneling time behavior in the region very close to $\alpha = 2$. This proximity provides a systematic and detailed understanding of the tunneling time behaviour as the system transitions from standard QM to SFQM. Furthermore, for $\alpha=1$, the SFSE reduces to a first-order differential equation where the time evolution is no longer directly governed by energy eigenstates as in the standard QM, which may result in non-local effects and anomalous diffusion. For $\alpha=1$, examining the tunneling time would be the natural direction to pursue for further research. 

\newpage
\noindent
{\it \bf{Acknowledgments}}:\\
\\
MU gratefully acknowledges Prof. P. Senthilkumaran, Head of the Optics and Photonics Centre (OPC) at the Indian Institute of Technology Delhi (IITD), New Delhi, for his invaluable encouragement and support in fostering research activities. MU also extends sincere appreciation to the OPC IITD, for fostering a supportive and stimulating research environment, which has been instrumental in advancing this work. VNS acknowledges the support from the School of Basic and Applied Sciences, IILM University, Greater Noida, for providing a conducive research environment. B.P.M. acknowledges the incentive research grant for faculty under the IoE Scheme (IoE/Incentive/2021-22/32253) of the Banaras Hindu University.%

\vspace{10pt}

\bibliographystyle{elsarticle-num}
\bibliography{References}
\end{document}